\documentclass{article}

\usepackage{arxiv}

\usepackage[utf8]{inputenc} % allow utf-8 input
\usepackage[T1]{fontenc}    % use 8-bit T1 fonts
\usepackage{hyperref}       % hyperlinks
\usepackage{url}            % simple URL typesetting
\usepackage{booktabs}       % professional-quality tables
\usepackage{amsfonts}       % blackboard math symbols
\usepackage{nicefrac}       % compact symbols for 1/2, etc.
\usepackage{microtype}      % microtypography
\usepackage{lipsum}

\usepackage{geometry}
\usepackage{cite}
\usepackage{float} 
\usepackage{amsmath,amssymb,amsfonts}
\usepackage{hyperref}
\usepackage{graphicx}
\usepackage{graphics}
\usepackage{textcomp}
\usepackage{multirow}
\usepackage{xcolor}
\usepackage{soul}
\usepackage{color}
\usepackage{hyperref}
\usepackage{algpseudocode}
\usepackage{algorithm}

\title{Evaluation of AI-Supported Input Methods in Augmented Reality Environment}

\author{ 
Akos Nagy\\
Department of Networks \& Digital Media \\
School of Computer Science \& Maths,\\
ECE, Kingston University\\
Kingston upon Thames, UK\\
\texttt{A.Nagy@kingston.ac.uk}\\
\And
Thomas Lagkas\\
Department of Computer Science\\
International Hellenic University Kavala Campus\\
Greece \& South-East European Research Centre\\
\texttt{tlagkas@ieee.org}\\
\And
Panagiotis Sarigiannidis\\
Department of Electrical \\and Computer Engineering  \\
University of Western Macedonia\\
Kozani,Greece \\
\texttt{psarigiannidis@uowm.gr} \\
\And
Vasileios Argyriou\\
Department of Networks \& Digital Media \\
School of Computer Science \& Maths, \\
ECE, Kingston University\\
Kingston upon Thames, UK \\
\texttt{Vasileios.Argyriou@kingston.ac.uk}\\
}

\begin{document}

\maketitle

\begin{abstract}
 Augmented Reality (AR) solutions are providing tools that could improve applications in the medical and industrial fields. Augmentation can provide additional information in training, visualization, and work scenarios, to increase efficiency, reliability, and safety, while improving communication with other devices and systems on the network. Unfortunately, tasks in these fields often require both hands to execute, reducing the variety of input methods suitable to control AR applications. People with certain physical disabilities, where they are not able to use their hands, are also negatively impacted when using these devices.
 The goal of this work is to provide novel hand-free interfacing methods, using AR technology, in association with AI support approaches to produce an improved Human-Computer interaction solution.
\end{abstract}


\section{Introduction}

Augmented Reality is a rapidly evolving technology that has the potential to transform a wide range of industries, from healthcare \cite{doi:10.2147/AMEP.S249891, dennler2021augmented,8613651} to manufacturing \cite{app11125592, XU2021530}. By overlaying digital information onto the physical world, provided by systems on a network \cite{app11041738}, AR offers a powerful tool for training, visualization, and work scenarios, enabling users to increase efficiency, reliability, and safety. However, one of the challenges facing AR is the limited range of input methods that are available for controlling AR applications, particularly in scenarios where both hands are required to perform a task. This limitation not only impacts the usability of AR but also has significant implications for people who, either require both hands for work processes, observed in industrial or medical fields, or with certain physical disabilities, who may not be able to use their hands to interact with these devices \cite{s22207719, RASHID2017248}.

A variety of input methods are available using AR technologies\cite{Kato2000, Billinghurst2001}, most of which face different limitations in working environments \cite{Pfeuffer2021, vertegaal2003attentive, 10.1007/978-3-030-85623-6_32}. Hand tracking allows the user to use their hand to interact with virtual objects or perform certain gestures to communicate with the device, which limits the use case when both hands are required while working. Additional peripherals can provide inputs, but these devices add additional cost to purchase, while operating them can require non-natural, or cause limited movement\cite{whitlock2018interacting, Muller2019, 6977392, Jacob1995}. To realise these AR interfaces and solutions, technologies based on computer vision, 3D reconstruction, scene analysis, motion estimation, and object detection are required \cite{ARGYRIOU2010887,ip-vis_20051073}. Freehand input control offers potential solutions to these challenges, allowing users to interact with AR applications without the need to use their hands. This technology enables users to control AR applications through natural body movements, such as head movements and voice commands \cite{Bates1995}, freeing up their hands for other tasks. Voice commands can be limited in certain environments, where the general background noise level is too high for reliable voice perception, \cite{10.4108/icst.intetain.2015.259629, app12031091}. 

As a viable interaction solution, this work focuses on improving head orientation-based input methods by evaluating head- and image-based input solutions assisted by Interpolation and Gravity-Map AI support approaches and compares them with classical input methods, such as mouse and gamepad. The main contributions of this paper are the following:
\begin{itemize}
    \item[a)] Introduce a novel AI-supported interfacing approach to improve user experience 
    \item[b)] Defined evaluation framework for comparing input methods, including task specification and evaluation metrics
    \item[c)] Evaluation of AI-support-assisted alternative input approaches and comparison with classical input methods
\end{itemize}

The paper is structured as follows: Section 2 provides a review off related research and methods. The proposed AI-enabled interaction methods for hands-free applications are detailed in section 3. Section 4 showcases the results using both the classic and the proposed interaction methods in a comparative study. Finally, the conclusion is presented in section 5.

\section{RELATED WORK}

Human-Computer Interaction (HCI) is a core element of AR use cases. Augmentation allows adaptive HCIs, that form to the human body, or allows itself to be an input device. Numerous studies contribute to finding the most optimal solutions in a wide range of scenarios. 

Frusto-Pascual \cite{10.1007/978-3-030-85623-6_29} examines different virtual keyboard positions and interaction feedback methods for character input in augmented reality. Two categories of keyboard position conditions are presented: viewpoint bound and non-dominant hand bound conditions. Viewpoint bound conditions are defined based on the position and orientation of the head-mounted display (HMD), while non-dominant hand bound conditions are defined based on the position and orientation of the non-dominant hand. Two forms of visual feedback and guidance for interaction with the keyboard are evaluated: Raycast and Glow. Raycast is a continuous guidance ray to the keyboard, while Glow is a visual representation of the dominant hand index fingertip. Both feedback methods are combined in the "Both combined" condition. 

 A system proposed by Chen \cite{Chen2020} is developed within a VR environment that allows freehand manipulation of 3D objects. The system includes a VR disambiguation menu with preview cubes, which vary in timing and modality based on the techniques. The three promising input modalities, explored for hands-free operation in VR are head gaze, speech, and foot. Also, the timing of disambiguation is explored. Although the methods are proposed related to VR environment, are applicable using AR devices.

 The foot-based method presents some limitations that may make it unsuitable, as mentioned before, for some users and certain environments. One significant limitation is that not everyone can use their feet effectively, such as individuals with certain disabilities or injuries that affect their lower extremities. Consequently, the foot-based method may not be an inclusive solution, and alternative selection methods should be considered to accommodate a broader range of users.

Moreover, the foot-based method requires additional equipment to track the user's foot movements accurately, which could increase the complexity of the system and may not be practical in industrial environments. For example, workers in manufacturing or construction sites may be wearing heavy-duty work boots, steel-toed shoes, or other safety equipment that could interfere with the accuracy of the foot-tracking.

Eye gaze-based interfaces for human-computer interaction are becoming increasingly popular and are being used in various sectors such as tele-operation \cite{Mahmud2020}. In tele-operation, the operator's eyes are occupied with monitoring the environment and their hands are busy controlling tasks. Eye gaze-based interfaces provide a natural user interface that is becoming increasingly popular and could become the interface of choice for many devices. Researchers have introduced various eye gaze tracking systems for tele-operation, including those for people with disabilities. A cost-effective and precise eye tracker has been developed, using an eye camera with two IR-emitting LEDs next to the camera lens \cite{Ho2014}. A gaze tracking system for tele-operation for people with physical disabilities was introduced, using an analog video-type camera to capture images and a gaze estimation algorithm to control the remote \cite{Yu2014}. An eye tracking-based mobile robot for tele-operation was also proposed, using a Microsoft Kinect 3D camera sensor for eye tracking. An eye gaze-based interface for a smart wheelchair was developed to assist people with locomotor disabilities. The system includes image processing for eye tracking and a module for controlling the movement of the wheelchair. It also includes added functionality such as sending messages through a smartphone and controlling electrical devices \cite{Gege2017}.

A study \cite{10.1007/978-3-030-85623-6_32} conducted in 2021 investigates the effectiveness of a gaze-adaptive user interface (UI) for Augmented Reality in comparison to an always-on interface, presented. Previous research has suggested that eye gaze could be a potential solution to the attention dilemma that arises when users must shift their focus between the real world and AR content \cite{Pfeuffer2021, vertegaal2003attentive}. The study presents the findings of an empirical investigation that evaluated the performance of gaze-adaptive UI in comparison to the always-on interface. The participants were asked to perform tasks that involved shifting their focus between the real world and virtual content. The results showed that most participants preferred the gaze-adaptive UI and found it less distracting. The gaze UI was faster, more intuitive, and perceived as easier to use when focusing on reality. When focusing on virtual content, the always-on interface was faster, but the user preferences were divided.

\begin{figure}[h]    
    \centering
    \includegraphics[scale=0.4]{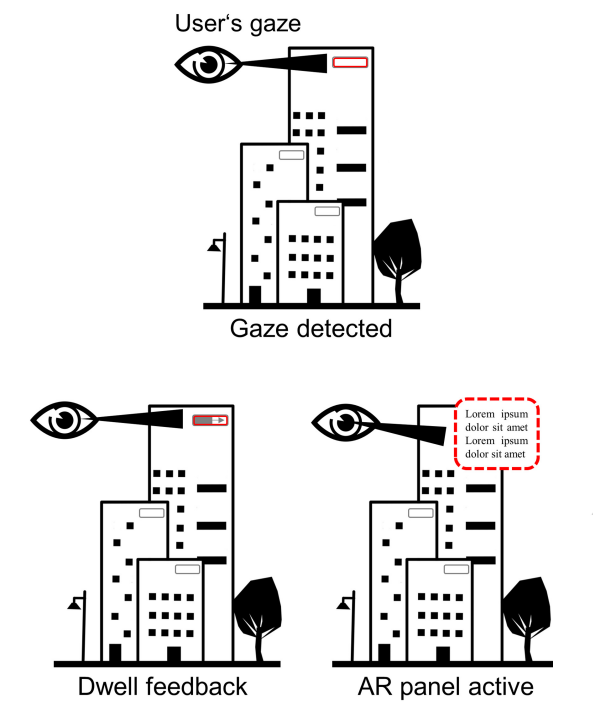}
    \caption{AR - AI Control Support Architecture \cite{10.1007/978-3-030-85623-6_32} }
    \label{fig:GazeBasedInformationRetrivalInAR}
\end{figure}

The proposed system, as presented with \autoref{fig:GazeBasedInformationRetrivalInAR}, provides additional information when the user gazes at a particular point, triggered by the temporal and spatial activation dimensions. Temporal activation determines how long a user needs to look at an object to trigger the AR panel, while spatial activation defines the size of the area a user needs to gaze at. The paper aims to formalize the properties of a gaze-adaptive AR system and better ground the design and study efforts.

\begin{figure*}[t]        
    \centerline{\includegraphics[width=0.8\textwidth]{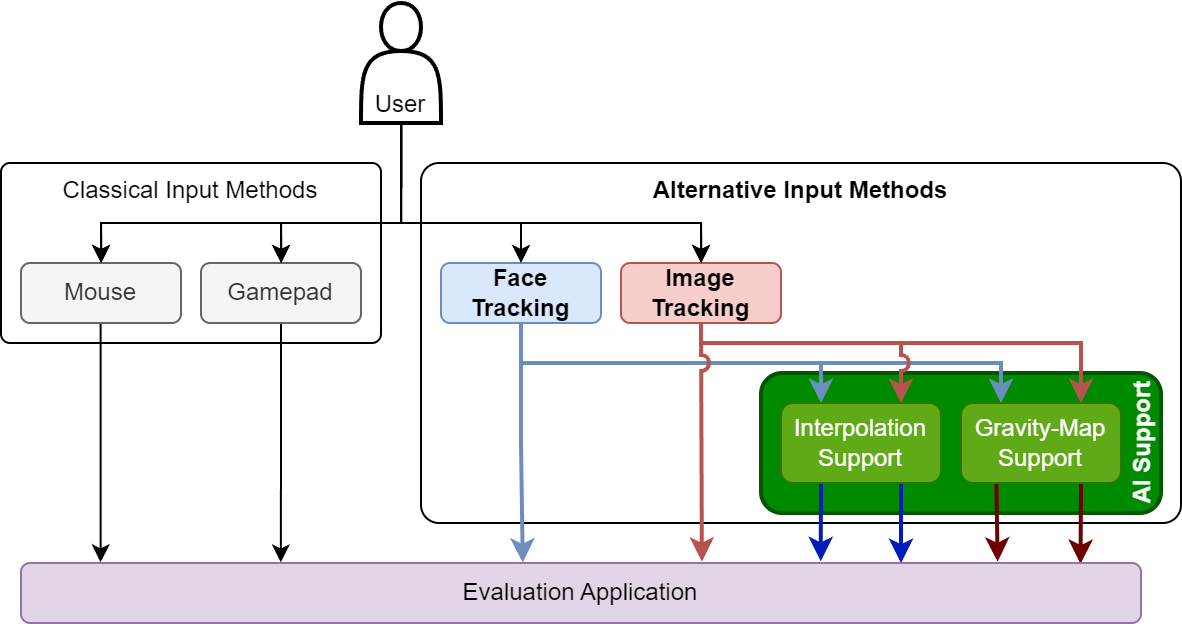}}
    \caption{AR - AI Control Support Architecture}
    \label{fig:AI_ARchitecture}
\end{figure*}

\section{METHODOLOGY}

Augmented Reality devices have the potential to revolutionize the way industrial workers receive or interact with information. AR devices can provide hands-free access to information and instructions and communication with network-connected devices, allowing workers to focus on their tasks while still having the necessary information at their disposal. This can be achieved through the use of head-mounted displays, smart glasses, or even smartphones. The AR technology overlays digital information in the real-world environment, providing workers with real-time access to important data such as instructions, diagrams, and technical specifications.

The present study focuses on Hand-free controls supported HCI in cases where using hands is not applicable. These cases include work environments where the person requires both hands to perform the tasks required by the processes, i.e.: industrial or medical fields. Other scenarios where due to an injury or a medical condition, where the person is not capable of using their hands also limit the usage of AR devices. Alternatively, a voice command could be used in these cases, but in certain environments, due to the high volume of background noise or the necessity of a quiet environment may not allow a such solution.

The following section describes a methodology of an AI-assisted architecture that supports hand-free control using AR technologies. The comparative study provides an evaluation of the performance and usability of the proposed hand-free control solutions compared to traditional input methods such as keyboard and mouse. The aim is to assess the potential of AR technologies for creating more intuitive and natural user interfaces for various applications.

While the mouse on personal computers, gamepads on console devices, and touch screens on mobile devices are part of the most common human-computer interfaces, in certain scenarios, these approaches might not be the most suitable. Physical limitations may be caused by the working environment, like industrial settings, where personnel requires both hands to perform the required actions, or due to disabilities or injuries, where the person's physical capabilities are impaired, these devices, due to their design, are not suitable for human-computer interactions.

The goal of the following proposed framework for HCI is to design a general pipeline architecture, demonstrated in \autoref{fig:AI_ARchitecture}, involving AI-assisted solutions, that allows the use of a variety of alternative input methods while minimizing the discrepancy in usability compared to classical input devices. In our study, we consider two input mechanisms as classical input devices, the mouse, most commonly used with personal computers, and gamepads, which are mainly used with console devices and dominantly for entertainment purposes. As alternative input methods, assisted image and face tracking-based solutions are proposed with applications in immersive environments. The face tracking method uses computer vision algorithms to determine the position and orientation of the user's head, and based on these parameters provides constantly input to the corresponding system, while the image tracking solution again based on computer vision methods, detects the selected image target. and constantly tracks its position and orientation in the 3D world coordinates.

\subsection{AI Support Approaches}

The proposed AI-assisted solutions, integrated into the architecture, are designed to reduce the difference between the classical input devices and the alternative input solutions or devices. Using the input vector these solutions will alter the data before providing it to the testing application. The current study considers three different solutions as part of the experiment:

\textbf{Interpolation} is a powerful technique in computer graphics and computer vision that can be used to estimate and derive values between known data points. In the context of hand-free control, the interpolation technique allows for the deviation of the input vector toward the intended target based on a specific set of limitations and criteria. The input vector's deviation is determined by considering the distance of the target from the input vector and the original orientation of the input vector.

This balance between forced, where the AI is influencing the input method, and free control, where there is no influence, is crucial in ensuring that the system is both responsive and intuitive to the user's input, while also maintaining a level of control over the target object. The technique allows for smooth and gradual adjustments to the input vector so that the target can be reached in an efficient and accurate manner. 

The Pseudo code for the implementation is as seen in \autoref{alg:InterpolationPrediction}.

\begin{algorithm}[h]
\scalebox{0.9}{
    \begin{minipage}{1\linewidth}
    \caption{InterpolationPrediction}\label{alg:InterpolationPrediction}
    \begin{algorithmic}[1]

    \Require{\\
        $start$: current coordinate,\\ 
        $moveVec$: current input vector ,\\ 
        $influence$, default(0.8f)\\ 
        $number$: number of iterations, how many possible future positions need to be predicted; default(1)
    }
    \Ensure{ $p$: an array of predicted influenced points}
    \Function{InterpolationPrediction}{$start$, $moveVec$, $target$, $influence = 0.8$, $number =  1$}
        \State $p \gets [start, start + moveVec]$ \\
        \For{$i \gets 0$ \textbf{to} $number$}
            \State $c \gets p[\operatorname{length}(p) - 2]$
            \State $tVec \gets target- c$ \\
            \If{$moveVec.\operatorname{mag}() == 0 \lor tVec.\operatorname{mag}() == 0$}
                \State append $c$ to $p$
                \State \textbf{continue}
            \EndIf \\
            \State $tVecNorm \gets \operatorname{normalize}(tVec)$
            \State $moveVecNorm \gets \operatorname{normalize}(moveVec)$ \\
            \State $mod \gets \operatorname{max}(tVecNorm.\operatorname{dot}(nVecNorm), influence) - influence$
            \State $mod \gets mod \times (1 / (1 - influence))$ \\
            \State $fVec \gets \operatorname{lerp}(moveVecNorm, tVecNorm, mod)$ 
            \State $modDist \gets \operatorname{min}(tVec.\text{mag}(), moveVec.\text{mag}())$
            \State $next \gets fVec \times modDist + c$    
            \State append $next$ to $p$
        \EndFor \\
        \State \Return elements of $p$ starting from index $1$
    \EndFunction
\end{algorithmic}
\end{minipage}
}
\end{algorithm}

The \textbf{Gravity-Map} approach solution for controlling input vectors is a more flexible approach compared to the Interpolation approach and can be used to control the deviation of the input vector toward multiple target objects. In this approach, the interface is divided into two areas, with one area being defined as the area of effect, and the other being the unaffected area. The area of effect is determined by the proximity of the target objects to the current focus position. When the cursor position is within the area of effect, the input vector is deviated toward the intended target object based on its distance and original orientation.

The Gravity-Map approach is different from other interpolation techniques in that it considers the relationship between the input vector and multiple target objects, rather than just one target object at a time. This allows for a more intuitive control experience, as the input vector is pulled towards the target objects that are closest and within reach. Furthermore, dividing the interface into two areas provides a balance between forced and free control. In the area of effect, the input vector is automatically directed toward the target objects, providing assistance for users. However, outside of this area, the input vector remains unaltered, giving users the freedom to move and control the cursor as they see fit.

The Pseudo code for the implementation is as seen in \autoref{alg:GravityMap}.

\begin{algorithm}[h]
\scalebox{0.9}{
    \begin{minipage}{1\linewidth}
\caption{Calculate influence vector at coordinate}
\label{alg:GravityMap}
\begin{algorithmic}[1]

\Require{\\targets (list of rectangles with $x,y$, width, height attributes),\\
$PX$ (x coordinate of a point), \\
$PY$ (y coordinate of a point), \\
$influenceDistance$ (the max distance a given target is allowed to influence the outcome)}
\Ensure{influence vector at point described by $PX$ and $PY$}
\Function{GravityMapInfluence}{$targets$, $PX$, $PY$, $influenceDistance$}
\State $retInfluence \gets (0, 0)$;

\For{$i \gets 0$ \textbf{to} $\text{length}(targets)$}

\State $closestPoint \gets$ calculate closest point within the target area to $(PX, PY)$;

\If{$(PX, PY)$ is within the target area}
\State \Return $(0, 0)$;
\EndIf

\If{distance between $closestPoint$ and $(PX, PY)$ is greater than $influenceDistance$}
\State \textbf{continue};
\EndIf

\State $direction \gets$ calculate direction vector towards $closestPoint$;
\State $distanceRatio \gets$ distance between $closestPoint$ and $(PX, PY)$ / $influenceDistance$;
\State $influence \gets direction \times (1 - distanceRatio^2)$;

\State $retInfluence \gets retInfluence + influence$;
\EndFor

\State \Return $\mathbf{retInfluence}$;
\EndFunction
\end{algorithmic}
\end{minipage}
}
\end{algorithm}

\subsection{Data-collection Methods}

\begin{figure}[h]    
    \centering
    \includegraphics[scale=0.5]{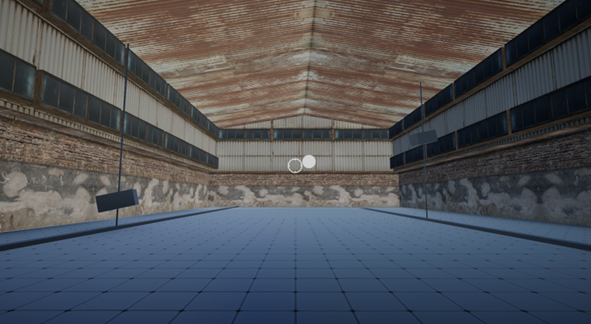}
    \caption{Evaluation application (in follow mode) screenshot with crosshair and target (white) }
    \label{fig:AppSample2}
\end{figure}

Data collection has been performed via a data evaluation application (\autoref{fig:AppSample2}), which encompasses three modes, each with its own task to perform and metrics to evaluate. The study involved 20 participants.

\textbf{Locate Mode}

In this mode, a series of static targets are displayed within the 3D environment. The participant's objective is to align the focus point, which is represented with a crosshair in the application, with the target. The task consists of multiple targets, each of which constitutes a separate subtask, organized in a sequential manner with only one target being visible at any given time. In this mode, we measure the average time it takes for a participant to reach the target position with the various input methods and solutions. Each target has a defined window of availability, beyond which it will be removed, and the corresponding subtask will be considered as failed.

\textbf{Select Mode}

In this mode, the participant's task is to precisely position the crosshair, which serves as the center of the screen, onto the static targets in a 3D environment. The evaluation consists of a series of subtasks, each of which involves positioning the crosshair onto a target and maintaining it there continuously for a set amount of time. The targets are presented in a sequential manner, with only one target being visible at a time.

The accuracy of the various input methods and solutions was assessed by conducting experiments and measuring the ability of the input method to stay within a designated region of the screen. The success rate of each input method was calculated based on the number of successfully completed subtasks, and the extra time spent on the target was recorded for subtasks that took longer than necessary to complete.

\textbf{Follow Mode}

In the Follow Mode, the participants are tasked with tracking a moving target, as opposed to a static target in the other modes. The targets move along a predefined path at predetermined speeds to control the experiment and determine the relationship between success and speed. The objective is to position the crosshair, which serves as the center of the screen, onto the target and follow it along the designated path between the start and end marks. This mode evaluates the precision of the input solutions in a dynamic target scenario, similar to the Select Mode.

\section{Results}

\subsection{Quantitative Metrics}

\begin{figure}[h]    
    \centering
    \includegraphics[scale=0.5]{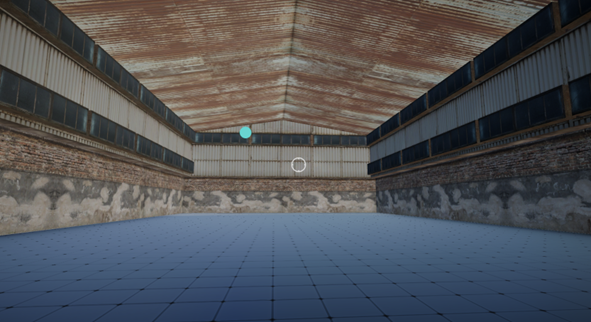}
    \caption{Evaluation application screenshot with crosshair and target (cyan) }
    \label{fig:AppSample}
\end{figure}

In this section, we will discuss the various metrics used for the Quantitative results. In the section, the focus point refers to the center of the screen, represented by a crosshair in the testing application, as displayed on \autoref{fig:AppSample}.

\textbf{Locate Mode}

\textit{Avg. Reach Time}: Let $S_{locate}$ be the set of sub-tasks and let $n$ be the total number of sub-tasks in $S_{locate}$. Let $S'_{locate}$ be the subset of $S_{locate}$ consisting of the sub-tasks that were completed successfully. A sub-task constitutes success if the focus point overlaps with the target object. For every sub-task, we record two timestamps, $TargetAppearedTimeStamp$, when the target object appeared on the game, and $TargetReachedTimeStamp$, when the focus point reached the target object. From these values, we calculate a $ReachTime$ for every sub-task, which is the difference between the previous two timestamps.

\begin{equation*}
    \text{Avg. Reach Time} = \frac{1}{n} \sum_{s \in S'_{locate}} s_{ReachTime}
\end{equation*}

\begin{equation*}
    s_{ReachTime} = s_{TargetAppearedTimeStamp} - s_{TargetReachedTimeStamp}
\end{equation*}

\textit{Target Reached \%}: Let $S_{locate}$ be the set of sub-tasks and let $n$ be the total number of sub-tasks in $S_{locate}$. Let $S'_{locate}$ be the subset of $S_{locate}$ consisting of the sub-tasks that were completed successfully. This formula calculates the average of successful sub-tasks by dividing the sum of all successful sub-tasks by the total number of sub-tasks in $S_{locate}$.

\begin{equation*}
    \text{Target Reached \%} = \frac{1}{n} \sum_{s \in S'_{locate}} s
\end{equation*}

\textbf{Select Mode}:

\textit{Avg. Extra Time Required}: Let $S_{Select}$ be the set of sub-tasks in Locate Mode and let $n$ be the total number of sub-tasks in $S_{Select}$. Let $S'_{Select}$ be the subset of $S_{Select}$ consisting of the sub-tasks that were completed successfully. A sub-task constitutes success if the focus point overlaps with the target object for a set $SelectTarget$ time.

\begin{equation*}\resizebox{0.5\textwidth}{!}{$\text{Avg. Extra Time Required} = \frac{1}{n} \sum_{s \in S'_{select}} (s_{T_{overlap}}-SelectTarget)$}
\end{equation*}

\textit{Target Selected \%}: Let $S_{Select}$ be the set of sub-tasks in Select Mode and let $n$ be the total number of sub-tasks in $S_{Select}$. Let $S'_{Select}$ be the subset of $S_{Select}$ consisting of the sub-tasks that were completed successfully. This formula calculates the average of successful sub-tasks by dividing the sum of all successful sub-tasks by the total number of sub-tasks in $S_{Select}$.

\begin{equation*}
    \text{Target Selected \%} = \frac{1}{n} \sum_{s \in S'_{select}} s
\end{equation*}
\hfill \break 
\textbf{Follow Mode}:

\textit{Avg. Follow \%}: Let $S_{Follow}$ be the set of sub-tasks in Follow Mode and let $n$ be the total number of sub-tasks in $S_{Follow}$. Let $S'_{Follow}$ be the subset of $S_{Follow}$ consisting of the sub-tasks that were completed successfully. This formula calculates the average percentage of overlap time during the successful sub-tasks by dividing the sum of all percentages of overlap time for sub-tasks by the total number of sub-tasks in $S'_{Follow}$. The percentage of overlap time for a sub-task is the overlap time divided by the fly-time of the sub-tasks. Fly-time is the total amount of time the target object is available on the screen, and overlap time is the time the focus point is overlapping with the target object during the sub-task.

\begin{equation*}
        \text{Avg. Follow \%} = \frac{1}{n} \sum_{s \in S'_{Follow}} (s_{OverlapTime} / s_{FlyTime})
\end{equation*}

\textit{Moving Target Touched \% }: Let $S_{Follow}$ be the set of sub-tasks in Follow Mode and let $n$ be the total number of sub-tasks in $S_{Follow}$. Let $S'_{Follow}$ be the subset of $S_{Follow}$ consisting of the sub-tasks that were completed successfully. This formula calculates the average of successful sub-tasks by dividing the sum of all successful sub-tasks by the total number of sub-tasks in $S_{Follow}$. A sub-task constitutes success if the focus point overlaps with the target object.

\begin{equation*}
        \text{Moving Target Touched \% } = \frac{1}{n} \sum_{s \in S'_{Follow}} s
\end{equation*}

\subsection{Data evaluation}

The evaluated data is separated into three tables, \autoref{table:ResultLocateMode}, \autoref{table:ResultSelectMode} 
 and \autoref{table:ResultFollowMode}, one for each of the data-collection modes, displaying the previously discussed metrics associated with the modes. Each table is also separated into two sections to showcase the differences between the input methods without AI support, as baseline values, and with AI support, to evaluate the efficiency of the support approaches. Each section has the best-performing values highlighted. 

The results show that the Mouse input method is the overall best-performing among the input methods evaluated in this study. Among the AI-supported input methods, the Head movement-based, Gravity-Map approach performed the best, in certain aspects even better than the Mouse input.

\textbf{Locate Mode}

While the Mouse input method performed 48\% better than the overall average in terms of Reach Time, it's only 18\% better than the Gravity-Map assisted Head input, which has gained a 42\% improvement compared to not having AI support. Image-based input was the worst-performing, not AI-supported method, and while the Gravity-Map AI-support approach slightly improves the method, using the Interpolation approach increased the Average Reach Time by 18\%. 

\begin{table}[H]
\centering
\caption{OUTCOMES FOR ALL THE INPUT METHODS IN LOCATE MODE}
\label{table:ResultLocateMode}
\resizebox{0.5\columnwidth}{!}{%
\begin{tabular}{c|r|cc}
\hline
& Input & Target Reached \% & Avg. Reach Time \\
\hline
\parbox[t]{2mm}{\multirow{4}{*}{\rotatebox[origin=c]{90}{No AI Support}}} 
& Mouse & \textbf{98.3\%} & \textbf{0.89s} \\
& Gamepad & 97.8\% & 1.34s \\
& Head & 96.0\% & 1.86s \\
& Image & 75.5\% & 2.17s \\
\hline 
\parbox[t]{2mm}{\multirow{4}{*}{\rotatebox[origin=c]{90}{AI Support}}} 
& Head - Interpolation & 96.2\% & 1.78s \\
& Head - Gravity-Map  & \textbf{99.78\%} & \textbf{1.08s} \\
& Image - Interpolation & 76.1\% & 2.56s \\
& Image - Gravity-Map & 83.6\% & 1.82s \\
\hline 
& Overall Average & 90.4\% & 1.69s \\
\hline 
\end{tabular}%
}
\end{table}
\newpage
\textbf{Select Mode}

The Mouse input method performed over 9X better than the overall average in terms of Extra Time Required, due to the fact that a mouse does not have continuous input like the alternative input methods. While using the alternative methods, there's a continuous movement, even in the target (head, image) is seemingly motionless. The Head-based input using the Gravity-Map approach performed over 4X better, and became the third fastest solution, than the Head-based input without AI support, which required 2X more time to perform the tasks than the overall average, being the overall slowest solution. In terms of the Select Mode, a higher Extra Time Required value indicates a less stable input solution when required to dwell on a fixed point.

\begin{table}[H]
\centering
\caption{OUTCOMES FOR ALL THE INPUT  METHODS IN SELECT MODE}
\label{table:ResultSelectMode}
\resizebox{0.5\columnwidth}{!}{%
\begin{tabular}{c|r|cc}
\hline
& Input & Target Selected \% & Avg. Extra Time Required \\
\hline
\parbox[t]{2mm}{\multirow{4}{*}{\rotatebox[origin=c]{90}{No AI Support}}} 
& Mouse & \textbf{100\%} & \textbf{0.042s} \\
& Gamepad & 95.8\% & 0.099s \\
& Head & 85.5\% & 0.779s \\
& Image & 70.5\% & 0.470s \\
\hline 
\parbox[t]{2mm}{\multirow{4}{*}{\rotatebox[origin=c]{90}{AI Support}}} 
& Head - Interpolation & 81.9\% & 0.659s \\
& Head - Gravity-Map  & \textbf{99.6\%} & \textbf{0.178s} \\
& Image - Interpolation & 72.6\% & 0.494s \\
& Image - Gravity-Map & 90.1\% & 0.343s \\
\hline 
& Overall Average & 87.0\% & 0.383s \\
\hline 
\end{tabular}%
}
\end{table}

\textbf{Follow Mode}

The worst-performing solution was the Image-based input assisted by the Interpolation AI-support, resulting in only 50\% Avg. Follow \% compared to the average and 20\%-25\%, compared to the best-performing solutions. The two best-performing solutions were the Mouse and Head-based input method, assisted by the Gravity-Map AI-support approach. Only these two solutions were able to perform better than the overall average, in terms of Avg. Follow \%, with the Mouse-based input method performing 16\% better than the second-best solution.

\begin{table}[H]
\centering
\caption{OUTCOMES FOR ALL THE INPUT METHODS IN FOLLOW MODE}
\label{table:ResultFollowMode}
\resizebox{0.5\columnwidth}{!}{%
\begin{tabular}{c|r|cc}
\hline
& Input & Moving Target Touched \% & Avg. Follow \% \\
\hline
\parbox[t]{2mm}{\multirow{4}{*}{\rotatebox[origin=c]{90}{No AI Support}}} 
& Mouse & \textbf{99.4\%} & \textbf{48.1\%} \\
& Gamepad & 99.1\% & 19.3\% \\
& Head & 83.3\% & 14.2\% \\
& Image & 75.5\% & 17.9\% \\
\hline 
\parbox[t]{2mm}{\multirow{4}{*}{\rotatebox[origin=c]{90}{AI Support}}} 
& Head - Interpolation & 81.5\% & 11.9\% \\
& Head - Gravity-Map  & \textbf{99.5\%} & \textbf{41.3\%} \\
& Image - Interpolation & 66.2\% & 11.1\% \\
& Image - Gravity-Map & 83.9\% & 14.9\% \\
\hline
& Overall Average & 86.0\% & 22.3\% \\
\hline 
\end{tabular}%
}
\end{table}

\textbf{Overview}

\begin{figure*}[h]        \centerline{\includegraphics[width=0.9\textwidth]{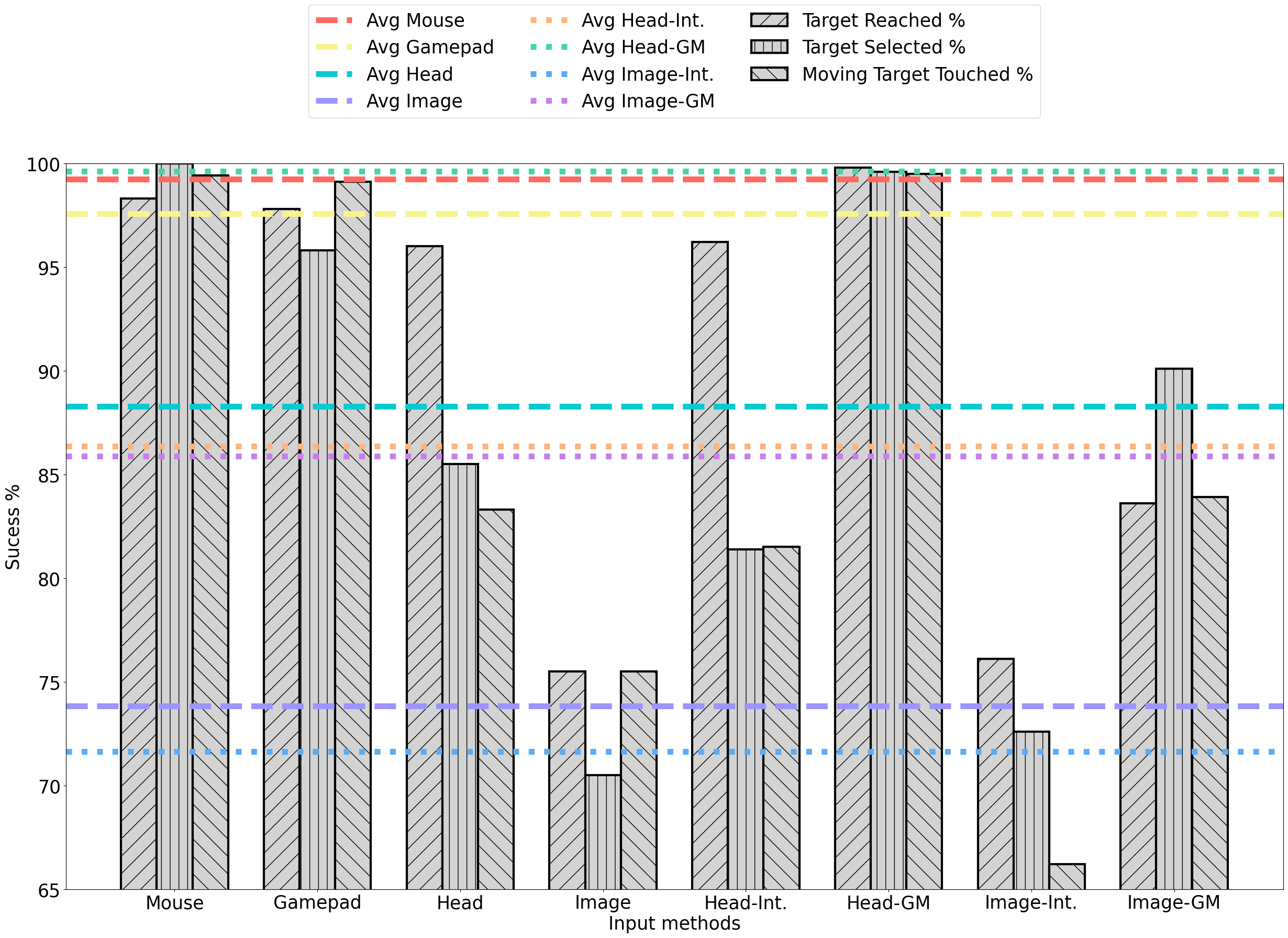}}
    \caption{Graphical overview of "Success \%" for each input-method and their relative average values}
    \label{fig:SuccessChart}
\end{figure*}

We consider Target Reached \%, Target Selected \% and Moving Target Touched \% as measures of success within their own data-collection mode, for all of these values relate to the ratio between successful and failed sub-tasks. \autoref{fig:SuccessChart} displays the relative success rate for each input method. Input methods are shown along the X-axis while the success rate is along the Y-axis. Vertical bars represent the success rate for each input method with different colors correlating to different data-collection modes. Horizontal lines represent the average success values for each of the input methods. We calculate this value by averaging Target Reached \%, Target Selected \%, and Moving Target Touched \% values. Each color represents each input method. Input methods without AI support are represented by dashed lines, while AI-supported methods are represented by dotted lines, to further increase readability. The diagram was also limited to the [65-100] range on the Y-axis, to emphasize the differences, that on a wider range would be not clear.

\section{CONCLUSIONS}

Considering that mouse (and keyboard) are the main input peripherals used in everyday life when controlling a personal computer, it was expected that the input method will produce the best performance. Based on the results we can conclude that using the Gravity-Map AI Support to aid alternative input methods, such as face tracking, significantly improves the viability of the method. The results show, that Head-tracking based input using the Gravity-Map AI support approach performs the closest to the mouse-based input. Taking into account that most of the participants only saw head tracking-based input methods first time during the testing process, we can presume that increased familiarity with the technology will lead to higher performance.

\section{ACKNOWLEDGMENT}

This project has received funding from the European Union’s Horizon Europe Research and Innovation programme under grant agreement No. 101070181.
\newpage
\bibliography{bib.bib}
\bibliographystyle{ieeetr}

\end{document}